# Ready for climate change?

# The importance of adaptive thermoregulatory flexibility for the Malagasy bat species *Triaenops menamena*

von
Sina Remmers
geboren am
10.04.1992

# Contents

# Abstract

The balance between energy intake and expenditure is essential and crucial for survival for all organisms. The energy management is closely linked to the ecology. Thus, changes in environmental conditions can be challenging, especially for the animals' physiology. Different strategies of thermoregulation have evolved and heterothermy seems to be the most efficient way for saving energy. Daily torpor, a temporally controlled reduction of the metabolic rate and body temperature, is one form of heterothermy and recent studies revealed that this physiological strategy is used by many tropical and subtropical species. Yet, little is known about torpor in bats and their intraspecific thermoregulatory flexibility. Therefore, three populations of the Malagasy bat species *Triaenops menamena* were investigated, to examine their metabolic rate (MR), skin temperature ($T_{skin}$) and related energy expenditure (DREE) during normothermic and torpid states in context of different microclimatic conditions. This study exposed significant physiological differences among these three populations along a gradient of fluctuation in environmental conditions. The greater the fluctuations in ambient temperature and humidity, the higher was the general resting metabolic rate and the rate of its reduction, but the lower was the torpid metabolic rate (TMR). Additionally, *T. menamena* expressed diverse patterns of torpor, e.g. hyperthermic daily torpor patterns with surprisingly high $T_{skin}$ (35.2°C) that even exceeded ambient temperature, while the MR was exceptionally low (0.06 ml $O_2$ $h^{-1}$ $g^{-}$). This species shows a highly adaptive flexibility in their physiology and are able to cope with unfavorable environmental conditions by using different strategies of thermoregulation and hypometabolism, which is beneficial regarding ongoing climatic changes. Further studies about seasonal variations are important and required to understand its broader physiological scope on an annual scale.

# Zusammenfassung

Das Gleichgewicht zwischen Energieaufnahme und -verbrauch ist für alle Organismen entscheidend und überlebenswichtig und dessen Regulierung ist eng mit der Ökologie verbunden. Daher können Veränderungen der Umweltbedingungen eine große Herausforderung darstellen, insbesondere für die Physiologie der Tiere. Es haben sich verschiedene Strategien zur Thermoregulation entwickelt, wobei Heterothermie der effizienteste Weg zur Energieeinsparung zu sein scheint. Der tägliche Torpor ist eine Form der Heterotherapie und äußert sich in einer zeitlich kontrollierten Reduzierung der Stoffwechselrate und Körpertemperatur. Bisherige Studien haben gezeigt, dass diese physiologische Strategie von vielen tropischen und subtropischen Arten angewendet wird. Über den täglichen Torpor bei Fledermäusen und deren intraspezifische thermoregulatorische Flexibilität ist jedoch wenig bekannt. Daher wurden drei Populationen der madagassischen Fledermausart *Triaenops menamena* untersucht, um mehr über deren Stoffwechselrate (MR), Hauttemperatur ($T_{skin}$) und den damit zusammenhängenden Energieaufwand (DREE) während normothermer und torpider Zustände in Bezug auf unterschiedliche mikroklimatische Bedingungen zu erfahren. Diese Studie zeigte signifikante physiologische Unterschiede zwischen diesen drei Populationen entlang eines Fluktuationsgradienten von Umweltbedingungen. Je größer die Schwankungen der Umgebungstemperatur und der Luftfeuchtigkeit waren, desto höher war die allgemeine Stoffwechselrate im Ruhezustand und dessen Reduktion, aber desto niedriger war die torpide Stoffwechselrate (TMR). Außerdem hat *T. menamena* verschiedene Muster des täglichen Torpors ausgeführt, z.B. einen hyperthermischen Torpor mit überraschend hoher $T_{skin}$ (35,2°C), die sogar die Umgebungstemperatur überstieg, während die MR außergewöhnlich niedrig war (0,06 ml $O_2$ $h^{-1}$ $g^{-}$). Diese Fledermausart zeigt eine hohe adaptive Flexibilität in ihrer Physiologie und ist in der Lage, mit ungünstigen Umweltbedingungen zurechtzukommen, indem sie verschiedene Strategien der Thermoregulation und des Hypometabolismus anwendet, was im Hinblick auf anhaltende klimatische Veränderungen von Vorteil ist. Weitere Studien über jahreszeitliche Variationen sind von großer Bedeutung, um mehr über die noch größere physiologische Flexibilität im Jahresverlauf zu erfahren.

# List of Tables and Figures

# Introduction

The balance between energy intake (foraging) and energy expenditure is essential and decisive for survival for all organisms. All animals consume energy for growth, reproduction, maintenance and it is indispensable for all other life-sustaining functions. Since animals interact with their abiotic and biotic environment and the success in foraging is limited to the available food sources of the habitat, their energy management is closely linked to the ecology. That is, most animals are restricted to specific environmental characteristics, which determine the distribution, abundance and behavior of every species. Changes in the conditions of these habitats can be challenging for every species and their physiology. Whether these habitat changes are induced by natural factors such as seasonal weather fluctuations or have an anthropogenic origin such as deforestation or climate change, they may cause higher energy expenditure (Heldmaier et al., 2013).

Furthermore, all endothermic mammals have the ability to maintain a constant and high body temperature over a wide range of ambient temperatures. However, small endotherms must produce higher amounts of endogenous heat than larger endotherms in order to avoid a reduction of body temperature while being exposed to colder weather. The reason for this is the ratio between body surface and area/volume, which increases with decreasing size. Yet, maintaining heat production and a high body temperature requires a steadily high food intake, which may be challenging due to fluctuating food availability (Geiser, 2004).

For these reasons, endotherms try to maximize energy intake and minimize expenditure using different approaches (Kronfeld-Schor & Dayan, 2013). Hence, different strategies of thermoregulation have evolved for saving energy, as not all mammals are homeothermic at all times. Besides strategies such as migration to more tolerable environments or energy reduction through behavioral and morphological adaptions, heterothermy and hypometabolism as adaptions to a changing microclimate, is one of the most effective and successful strategies to survive during periods of energy shortage (Kurta & Fujita, 1988).

In this regard, torpor is one form of heterothermy and is characterized by a temporary, controlled reduction of the metabolic rate (MR), body temperature and other physiological functions for decreasing energy and water needs, as a response to decreasing ambient temperature (Geiser, 2004). Torpor ensures maintaining the energy balance, even in unfavorable seasons and/or environments (Wang, 2000). Yet, each species has its specific thermoneutral zone (TNZ) – the range of ambient temperature in which the individual does not have to spend energy for maintaining a high body temperature. This range of the thermoneutral zone may not only change throughout the year (seasonal fluctuations) but also during different states of torpor (i.e. hibernation; Warnecke, 2017). During the state of torpor, the body temperature falls from normothermic degrees from ~32 - 42°C to degrees between -3 to < 30°C and the torpid metabolic rate (TMR) shows a 5-30% reduction of the basal metabolic rate (BMR) (Geiser & Ruf, 1995). The BMR is defined by the amount of energy required only for vital organs and is measured while the organism is at rest (not sleep) in its TNZ and when no exercise, reproduction, digestion or other activities take place (Mozo, 2019). Hence, the torpid state is highly effective, according to a possible metabolic rate reduction to less than 1% of the normothermic resting metabolic rate (RMR). The RMR is used more often than the BMR as a reference point for measuring an animal's physiology and its reduction under variable environmental conditions. This is due to the fact, that it is difficult to ensure that no higher variations in ambient temperatures and no digestion or movements take place during measurement periods, particularly during field measurements (Geiser, 2004).

There are different patterns of torpor used in different contexts and environments. Hibernators perform a prolonged torpor (i.e. hibernation), which usually lasts from autumn to spring. However, hibernators do not remain torpid for the whole season but exhibit torpor bouts for several days or weeks interrupted by periodic rewarming and normothermic resting periods. Many of them have to rely on stored fat or food as an energy source in winter (Geiser, 2004). Daily torpor is the other pattern, which is commonly used by many mammals and birds. Torpor bouts lasts only for hours in daily heterotherms (< 24 h) and are less seasonal (Geiser & Ruf, 1995). The daily torpor is usually interrupted by daily foraging/feeding and for nocturnal animals it usually takes place in later stages of the night and the early morning hours (Brigham et al., 2000; Geiser et al., 2000). Although daily torpor can occur throughout the year, its occurrence often increases in the colder and drier seasons. Even when food availability appears favorable, daily torpor is used regularly to balance energy budgets and not only as a response to acute energy shortage (Geiser, 2003). The body temperature reduction during daily torpor is not as deep as in hibernators – it usually falls to nearly 18° C in small mammals but can also be reduced below 10° C (Brigham et al., 2000). The metabolic reduction during daily torpor is about to 10 – 20% of the normothermic RMR and the overall daily energy expenditure is reduced to 50 – 90%, compared to days when no daily torpor is performed (Holloway & Geiser, 1995). Even though torpor and hibernation are mostly known for animals living in arctic regions, there are recent examples from various species from the tropics and subtropics, especially in Madagascar, where metabolic reduction is used to avoid environmental strains in arid, hot and unpredictable environments (Dausmann et al., 2004; Nowack et al., 2010; Ruf et al., 2015; Lovegrove & Génin, 2008; Kobbe et al., 2011). It is assumed, that there are much more functions of torpor in tropical regions (e.g. water conservation, removal of parasites, predator avoidance, longevity, etc.) than for the previously known arctic torpor (Geiser & Brigham, 2012; Nowack et al., 2017). In drier (sub-) tropical regions, the torpor as a strategy for water savings and conservation might be more important than a shortage in food availability, since all reductions in MR, activity and skin temperature lead to a diminution in water loss (Bondarenco et al., 2014; Cooper et al., 2005; Schmid & Speakman, 2009).

Madagascar is unique in terms of its environment and is globally known for its "biodiversity hotspot" with its rich and diverse endemic biota (Ganzhorn et al., 2001; Myers et al., 2000). In fact, evolutionary it has been an isolated island throughout history due to its separation from continental Africa 165 million years ago (Mya) and India 88 Mya. Furthermore, geographical reconstructions depict a northward drift starting around 45 Mya, which has led to its current geographical position, north of 30° southern latitude, and thus has been strongly influenced by the southeast trade winds ever since (Samonds, et al., 2012). Consequently, it is a place that is extremely affected by irregular climate events like droughts, cyclones or ocean current fluctuations, which have an impact on the ecosystem every year. It is characterized by an extreme diverse climate and therefore consists of a high amount of different habitat types (Dewar & Richard, 2007). Yet, the island suffers from extreme habitat loss (>90% of native habitat) and degradation, especially in areas with low topographic relief like the littoral and deciduous forests at the coast sites since these are more accessible (Myers et al., 2000; Green & Sussman, 1990). For this reason, endemic species from Madagascar are among the world's most endangered animals (Ganzhorn et al., 2001).

There are at least 46 different bat species documented in Madagascar's various and partially exceedingly challenging microhabitats. Yet, little is known about their habitat requirements and ecology, as well as their physiological demands or behavior (Lebarbenchon, et al., 2017; Goodman, 2011). This study concentrates on the species *Triaenops menamena*, which inhabits the drier regions and is endemic to Madagascar. They are mostly found in dry forests, just like in the south-western and western parts of Madagascar. These deciduous dry forests are less dense and have a more open canopy than humid forests, but they still exhibit a great floral diversity and species richness (Jenkins, 1987). Furthermore, the climate in these drier regions is characterized by a hot rainy season from November to March and a colder dry season between May and September. The temperature varies annually between 6 and 45°C and the precipitation is highly seasonal between 300 and 800 mm each year, which is concentrated on the rainy season and is spatially and temporally unpredictable (Kobbe et al., 2011; Lewis & Banar-Martin, 2012).

This study examines three *T. menamena* populations, two in the Tsimanampetsotse National Park (SW) and one in Kirindy forest (W) along a gradient of microclimate fluctuation. *T. menamena* is an insectivorous bat species, that prefers roosting in large colonies inside caves or in smaller numbers inside tree holes in forests when caves are absent (Goodman, 2011). In Tsimanampetsotse National Park, which is located on a calcareous plateau covered in dense spiny forest and includes numerous underground caves and stream systems, one population roosts in a cave with constant microclimatic conditions, whereas the other population roosts in a sink hole, which is more exposed to the daily fluctuations in ambient temperature and humidity (Reher et al., 2019). The population in Kirindy forest, which is part of the largest contiguous tracts of tropical dry forest in the country, roosts in tree holes and thus is highly affected by ambient conditions and weather fluctuations (Whitehurst et al., 2009). In this regard, the "Arrhenius effect" describes the relationship between ambient temperature and the metabolic rate and is directly linked to global warming through an increase in average global air temperatures and frequency and intensity of extreme climate (Lovegrove, et al., 2014). This Arrhenius effect could affect the metabolic rate and behavioral patterns of the three populations differently, as diverse microclimatic conditions may lead to intraspecific flexibility and adaptability (Dillon et al., 2010).

Bats in general include many heterothermic species, but there are only a few studies on heterothermy and hypometabolism in free-ranging tropical bats in their natural environment (Reher et al., 2018; Liu & Karasov, 2011; Turbill et al., 2003). For other species, recent studies revealed new physiological insights for animals living in tropical regions (Ruf & Geiser, 2015; Nowack et al., 2010). That is, especially different forms of torpor were uncovered, including patterns of metabolic reduction without any decrease in body temperature or its restriction to ambient temperatures. For example, the mouse lemur species *Microcebus griseorufus*, that also inhabits the dry regions of Madagascar, stayed in a torpid state even at high ambient temperatures and a corresponding body temperature of 37°C (Kobbe et al., 2011). Other studies disclosed hypometabolism at high ambient temperatures, with a torpid body temperature even above the normothermic temperatures (Lovegrove et al., 2014; Reher et al., 2018). Yet, another study of heterothermy in tropical mammals exposed patterns of thermoregulation, that are difficult to classify and assign to specific inhibition patterns like daily torpor or hibernation. The lesser Hedgehog Tenrecs *Echinops telfairi* were heterothermic most of the measurement time, even though they are definitely able to heat endogenously. It appeared to be more favorable to align with ambient temperatures and only be normothermic for short periods (McKechnie & Mzilikazi, 2011).

Considering bats, as small endotherms, have twice the metabolic rate than terrestrial mammals of the same size due to their capability of flying and their orientation through ultrasonic echolocation, they need to spend even more energy for performing daily activities (Thomas, 1975; Neuweiler, 2000). Additionally, heterothermy results in great ranges of body temperature and the largest range of skin temperature in mammals was in fact documented for the Australian desert bat *Mormopterus petersi*, with a skin temperature of up to 45.8°C during a heat wave and a decreased skin temperature of 3.3°C during winter season (Bondarenco et al., 2014). New studies from bats in Madagascar showed the variety in torpor patterns within one species. The species *Macronycteris commersoni* did not only show a form of "hot torpor" (state of hypometabolism at high ambient and skin temperature) with an exceptionally low metabolic rate (0.13 ml $O_2$ $h^{-1}$ $g^{-1}$), but also a yet unknown pattern of very short torpor bouts (on average 20 minutes). These results accentuate the capability of alternating fast between a normothermic and torpid metabolic rate and the animals' physiological scopes and flexibility (Reher et al., 2018).

The purpose of this study was to reveal yet unknown information about the physiology of tropical bats and their adaptions to unfavorable environmental conditions. Precisely, the investigation of the metabolic rate of *Triaenops menamena* through respiratory measurements will help to understand their physiological capabilities and intraspecific flexibility by using daily torpor to minimize energy expenditure. Environmental conditions, skin temperature and oxygen consumption were measured and compared between different microhabitats in order to answer following hypotheses:

(1) Bats roosting in more fluctuating microhabitats expend less energy in general, due to increased occurrence of daily torpor and a greater reduction in metabolic rate and skin temperature. A correlation between the bats body mass or condition and its oxygen consumption and skin temperature can be assumed.

(2) The time of entry into a torpid state is more synchronized in the forest habitat than in the caves, given the direct influences of the circadian cycle and environmental conditions. Moreover, the torpor duration is longer in the less buffered habitats, because of generally higher ambient fluctuations.

(3) Torpor patterns differ between each study site. The course of the metabolic rate and skin temperature each day is rather individually than species specific, assuming an intraspecific flexibility and adaptability regarding changes in ambient conditions.

For developing conservation plans to ensure a long-term persistence of endangered animals, it is essential to understand the different species' ecophysiological scopes and demands on their environment. This is of paramount importance for animals in (sub-)tropical regions, where they often already live at their physiological limits and the effects of global warming have a greater extent than in moderate temperature regions (Huey et al., 2012). It is unlikely that the current biodiversity will be maintained in the same microclimate habitats in the future and therefore, it is extremely important to understand more about the adaptive potential and physiological flexibility as a response to habitat changes (Green & Sussman, 1990; Ganzhorn et al., 2001).

# Materials and Methods

## *Study area*

This study was conducted in the Tsimanampetsotse National Park in south west Madagascar (Fig. 1). This region is one of the driest areas of the island, with an annual rainfall between 300 and 600 mm, which is mainly confined to the rainy season from November to March. However, the rain falls highly unpredictable with possible complete rain absence for several years (Rasoloariniaina et al., 2015). Our study was conducted in June and July 2019, during the colder dry season (May to September) with ambient temperatures from 6°C at night and up to 38°C during the day. Food and water availability are also influenced by the seasons and much lower during the dry season (Kobbe et al., 2011). The Park is located at the Mahafaly plateau, a calcareous plateau which is mainly covered in dense spiny forest and contains large numbers of caves and sinkholes, of which many contain water and underground steam systems (Rasoloariniaina et al. , 2015; Dobrilla, 2013).

Bats were trapped at two different caves in Tsimanampetsotse National Park. Andranolovy Cave (24.04585° S / 043.75396° E) is one of them and consists of several connected chambers underground that are partially flooded. The cave is well buffered with high constant temperature and relative humidity (RH) year-round. The smaller Vintany cave (24.04383° S / 043.75519° E) on the other hand is a sinkhole and therefore more affected by ambient microclimatic conditions. Both caves are only 270 m apart but display highly diverging environmental conditions and habitat features.

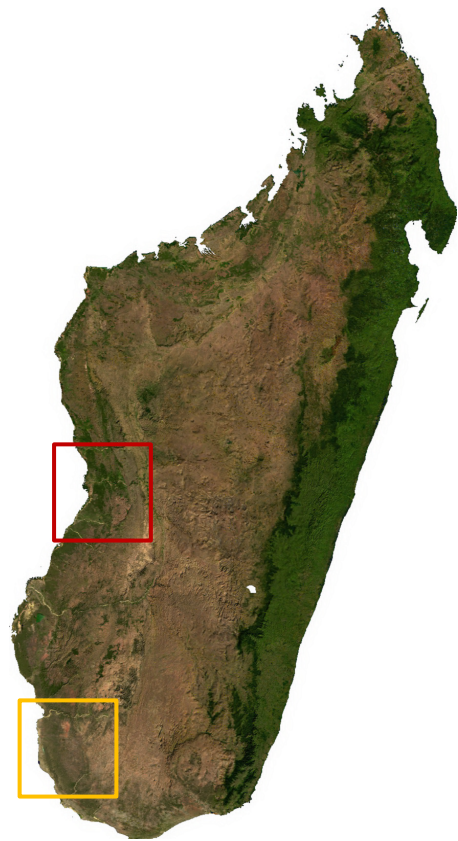

*Figure 1: Map of Madagascar with both study areas (see below: additional data) framed in red (Kirindy forest) and yellow (Tsimanampetsotse National Park) in the drier western part of the island (Map Maker Ltd, 2007)*

*Study species*

While gathering data during the fieldtrip, this study was focused on the species *Triaenops menamena* (Chiroptera: Rhinonycteridae), the Rufous Trident Bat. It was formerly known as *Triaenops rufus,* but new analysis and comparisons lead to a reclassification of this genus (Goodman & Ranivo, 2009). *T. menamena* is a medium-sized bat species with very variable fur coloration (reddish brown to gray); body mass averagely ranges from 6 to 16 g and a total forearm length from 40 - 55 mm on average (Peterson et al., 1995; Ranivo & Goodman, 2006).

*Triaenops menamena* is endemic to Madagascar where it is widespread (Ranivo & Goodman, 2006; Monadjem, et al., 2017). It is mostly common along the west coast, especially in the dry regions with spiny forests like in the southwest, but it can also be found in deciduous forests and the rainforest in the northeast (Goodman et al., 2005). One of its characteristics is to tolerate certain scopes of habitat transformation and is therefore also associated with forests where degradation is in progress (Russell et al., 2007).

The Rufous Trident Bat normally roosts in caves, often in large colonies, but can also be found roosting inside hollow trees in denser forests (Racey et al., 2009). Moreover, it is insectivore and mainly feeds on Lepidopterans, yet Coleopterans, Hemiptera and other insects are also components of its diet to cope with resource fluctuations throughout the seasons (Rakotoarivelo et al., 2007).

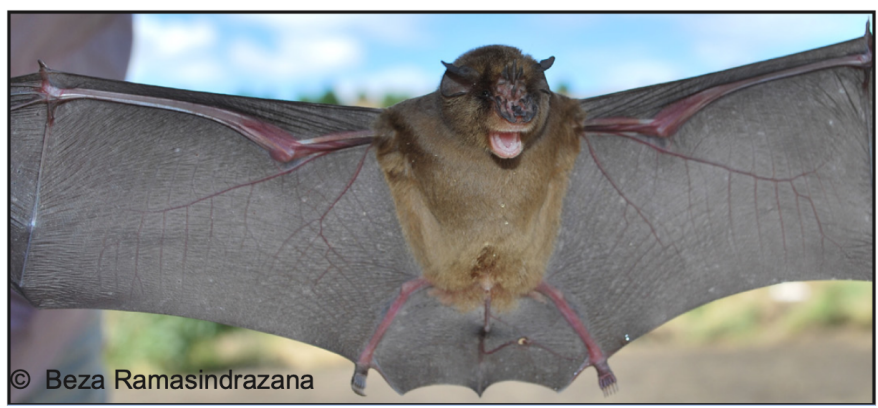

*Figure 2: Picture of the study species Triaenops menamena. (picture copyrighted by Beza Ramasindrazana) (Ramasindrazana & Goodman, 2014)*

## *Trapping and handling*

Bats were captured with a harp trap which was placed at the entrance of Andranolovy cave and the main corridor at Vintany which the bats used for leaving the sinkhole. The trap was opened about half an hour before sunset at 17:30, to be sure to capture them while leaving for foraging and left it open for three hours while checking on it every 15 minutes. A maximum of two adult bats was trapped per night due to our capacities for the respirometry measurements. Trapping was done for five consecutive days at one cave and then alternately changed the location to the other cave, to prevent for habituation and avoidance of the position of the trap. Thus, half the respirometry measurements were done for each site (started at Andranolovy cave and switched to Vintany cave after a week), before starting the second half of the measurement runs (again starting at Andranolovy cave). This setup also ensured to not having potential timing bias due to progressing season.

After transporting them from the harp trap to cloth bags sex, body mass (BM) and forearm length (FL) of each bat were determined and they were marked with a wing membrane tattoo of an individual 3-digit number. Therefore, local anesthesia (EMLA, AstraZeneca, Wedel, Germany) was applied before marking them with non-toxic ink (Hauptner-Herberholz, Solingen, Germany).

To register the skin temperature ($T_{Skin}$) during the respirometry measurements, temperature sensitive radio transmitter (0.45 g, Biotrack, Wareham, UK) were used. Hence, a small patch of fur was removed between the shoulder blades using a razor blade, then the area was dried and cleaned before attaching a transmitter with a medical latex adhesive (SAUER-Hautkleber, Lobbach, Germany). The temperature sensitive radio transmitters' weights were below recommended maxima for bats for load carrying capacity regarding radio telemetry measurements ("5%-rule"; Aldridge & Brigham, 1988). External transmitters were chosen over internally implanted radio transmitters, because in small mammals $T_{skin}$ provides an accurate reflection of the body's core temperature ($T_b$) and these transmitters do not require surgery, which could obstruct normal thermoregulation, and therefore might jeopardize the results (Audet & Thomas, 1996; Dausmann, 2005; Willis & Brigham, 2003).

*Metabolic rate and temperature measurements*

With the aim of determining the individual's energy expenditure and metabolic rate, the oxygen consumption was measured with an open flow-through respirometry system in pull mode. Therefore, bats were transferred into a 2L plastic metabolic chamber which was darkened from the outside and provided with a net for the bats to hang on. It also contained a Hygrochron iButton for logging the temperature and RH inside the metabolic measurement box (Hygrochron iButtons, Maxim integrated, San Jose, USA). Subsequently, the boxes were brought back inside the caves to perform all measurements in the bats natural roosting surrounding. In Andranolovy cave, *T. menamena* roosts in large species assemblages with *Macronycteris commersoni* and *Paratriaenops furculus* in the most buffered and flooded chamber of the cave as their preferred roosting chamber (Reher et al., 2019). For this reason, all respirometry measurements were implemented in an adjacent cave chamber, as this avoided unnecessary disturbances of the colony caused by frequent controls of the system, particularly for *M. commersoni* which is hibernating during the colder dry season. Even though the adjoining chamber was marginally colder (mean $\Delta T_{cave2\text{-}3}$ = ~3.9°C), the ambient conditions were stable during the whole study (daily temperature fluctuation < 1°C). It is also the main roosting chamber of the species *Miniopterus mahafaliensis*, which is as well active year-round due to similar diet preferences as *T. menamena* (Reher et al., 2019; Russell et al., 2007).

The bats metabolic rate was measured as the rate of oxygen consumption with a portable differential oxygen analyzer (OxBox; designed and constructed by T. Ruf and T. Paumann, University of Veterinary Medicine Vienna, Austria) which was connected to the metabolic chamber via gas-tight tubing (Tygon, Saint-Gobain COOP, Courbevoie, France) (Fig. 3). The connecting points were sealed supplementary with a two-component epoxy plasticine (UHU, Bolton Adhesives Rotterdam, Netherlands). The OxBox runs on a standard 12V car battery and contains electrochemical fuel-cell oxygen sensors (7OX-V CiTicel, Bieler + Lang, Achern, Germany). These $O_2$ sensors were calibrated in the laboratory right before and after the field trip using a gas-mixing pump which generated calibration gases by adding 0, 3, and 5% $N_2$ to the air, respectively (2KM300/a; Wösthoff Messtechnik GmbH, Bochum, Germany). Thereafter, a calibration value corresponding to an oxygen reduction of 1% was calculated using a linear regression.

Consequently, the sample air passed the metabolic chamber with a constant flow rate set at about 50 L/h using a flow pump, which was integrated to the open circuit respirometry system (Fig. 3). Before entering the gas analyzer in which the oxygen content was measured every 10s for 55m, the sample air was dried and filtered with silica gel. To control for any drift of the oxygen sensor, reference air from the surrounding of the metabolic chamber was analyzed for 5 minutes once per hour. All the data from the sample air and reference air were then stored on a SD memory card.

Measurements started right after capture and data was collected for about 24 hours on average, depending on the success of capture and the rapidity of handling the bats before transferring them into the measurement box. After each respirometry run, bats were weighed again for calculating mean body mass, before they were released at point of capture for the beginning of their usual active and forage phase.

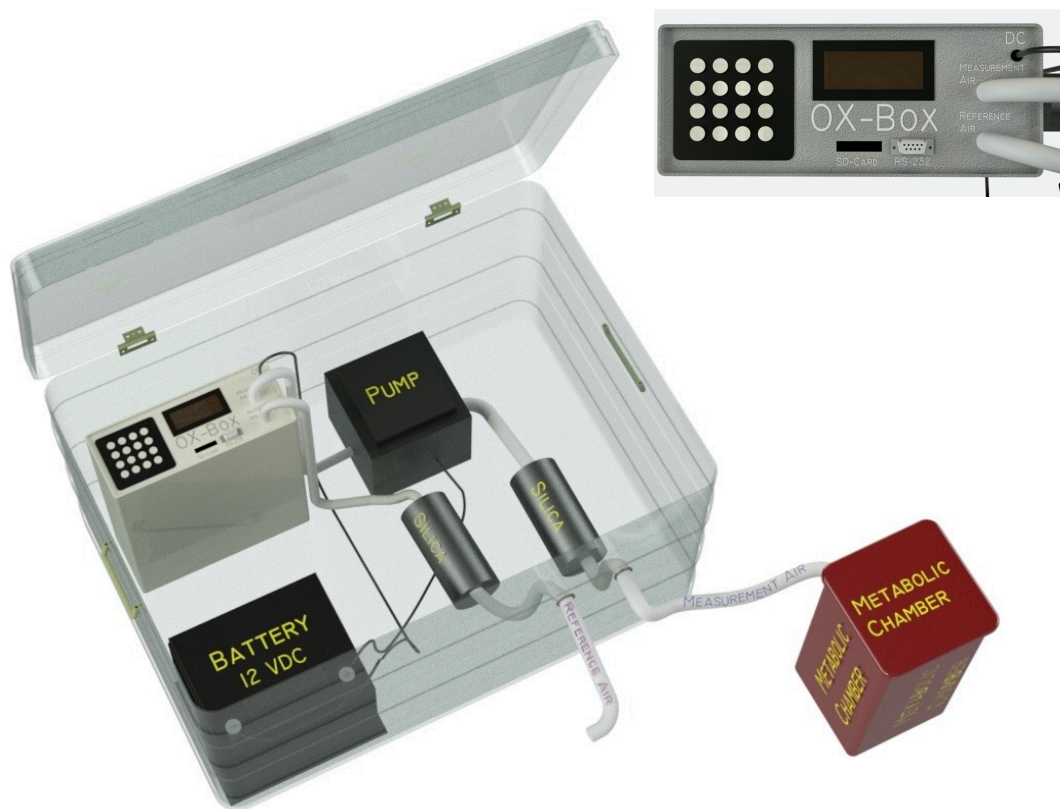

*Figure 3: Simplified visualization of the experimental setup of the oxygen consumption measurements. The OxBox was connected to the metabolic chamber (red) via gas-tight tubing for the measurement air and the reference air, with interposed pump and silica gel containers. It was also connected to a 12V battery and was equipped with a slot for a SD memory card (see enlarged picture on the upper right). During measurements, all components were stored in a metal box to protect the electronic devices from external damages and humidity. (Illustration created with AutoCAD (Autodesk Inc., 2019)*

The iButton inside the measurement box collected data ($T_{box}$ and $RH_{box}$) every 5 minutes, just like the iButton which was placed next to the metabolic chamber to record the ambient temperature and humidity ($T_{ambient}$ and $RH_{ambient}$). The temperature sensitive radio transmitters that were attached to the bats for recording $T_{Skin}$ collected data during the periods of the oxygen consumption measurements (see below) in a 5-minute interval using a remote/receiver logger (DataSika SRX-800-D, Biotrack, Wareham, UK). These transmitters were calibrated in advance in a water bath at 3-45°C against a precision thermometer (national standard). The radio signal and pulse period were monitored using a TRX-1000S Receiver and a stopwatch, after the transmitters remained in the water bath for 30 minutes at 3, 10, 17, 24, 31, 38 and 45°C to assure equilibration (Williams et al., 2009). To determine the relationship between pulse period and temperature, a least-squares regression was used to derive a regression equation that was later used for $T_{Skin}$ calculations ($R^2 > 0.99$ for all transmitters).

Additionally, long-term temperature and RH humidity data was recorded in the different cave chambers as well as outside the caves to account for the surrounding microclimatic conditions in the bats' habitat (Hygrochron iButtons, Maxim integrated, San Jose, USA). For this reason, one logger was placed near the cave entrances ($T_{envir}$ and $RH_{envir}$) and 3 loggers in several chambers inside Andranolovy cave; one in the main entrance chamber ($T_{cave1}$ and $RH_{cave1}$), one in the preferred roosting chamber ($T_{cave3}$ and $RH_{cave3}$) and one in the chamber where the respirometry measurements took place ($T_{cave2}$ and $RH_{cave2}$). These iButtons were affixed in about 1.5 m height and recorded data once per hour for 47 consecutive days (04.06.19-20.07.19).

*Additional data*

Regarding the wide distribution of *T. menamena* and to account for the larger scope of their thermoregulatory flexibility in different environmental conditions, additional data from Kirindy forest was included. This dataset was recorded in August 2018. The two study sites are both located in the drier western part of Madagascar (Fig. 1) but differ considerably in their microhabitats. Kirindy forest (44°39' E, 20°03' S) is a private park which is located around 50 km north-east of the town Morondava, with ranging temperatures from below 10°C at night to over 30°C during the day and almost no precipitation during the dry season (April – October) (Sorg et al., 2003; Dausmann et al., 2009). The park consists of a primary dry deciduous forest. Due to the absence of caves, the occurring bats roosting in hollow trees or comparable nests inside the forest under fluctuating ambient conditions. Long-term temperature and RH data was measured for 56 days in the dry season in 2018 (Hygrochron iButtons, Maxim integrated, San Jose, USA). Bats were trapped with mist nests along transects inside the forests. Handling procedure and the experimental setup were the same as in Tsimanampetsotse National Park, except for the position of the metabolic chamber and the measurement setup. The measurements took place in the open vegetation in the forest, considering the bats natural habitat and roost selection in Kirindy forest.

In total, data of 39 individuals was compared from three different microhabitats along a "fluctuation gradient", with 16 bats from Andranolovy as the least fluctuating cave, 12 bats from Vintany as a moderate fluctuating cave and 11 bats from Kirindy forest as the most fluctuating habitat.

*Data analyses*

The output values from the OxBoxes were corrected for drift manually using the program Clampfit v10.7.0.3 (Molecular Devices, Sunnyvale, USA), with the hourly reference air measurements serving as baseline values ($\Delta O_2$). Thereafter the differences in oxygen concentration between ambient reference air and sample air was calculated as follows:

$$\Delta volO_2 = \Delta O_2 / calib$$

with $\Delta volO_2$ = oxygen difference between sample air and reference air in %; $\Delta O_2$ = corrected output value from the OxBox; *calib* = calibration value derived from the calibration regression reflecting an oxygen reduction of 1%

The oxygen consumption ($\dot{V}O_2$) was calculated as ml $O_2$ per hour according to Lighton (2008), using the equation for excurrent flow measurements in flow-through respirometry with only dried air entering the analyzer and without removing $CO_2$ prior the measurements:

$$\dot{V}O_2 = \frac{FR_e \times F_iO_2 - F'_eO_2}{(1 - F_iO_2) \times 1 - RQ}$$

with $FR_e$ = overall excurrent flow rate; $F_iO_2 - F'_eO_2$ = difference in fractional concentration of incurrent and excurrent oxygen in the metabolic chamber (here: $F_iO_2 - F'_eO_2 = \Delta volO_2/100$); $F_iO_2$ = fractional concentration of incurrent oxygen (for outdoor air: $F_iO_2$ = 0.2095); $RQ$ = respiratory quotient (assuming an average respiratory quotient of 0.85 reflecting a metabolic combustion of 50% fat and 50% carbohydrate) (Lighton, 2008; Dausmann et al., 2009).

The daily resting energy expenditure (DREE) was calculated by including the oxycaloric equivalent, to account for the enthalpy change of the oxidative catabolic reactions per amount of oxygen respired (Gnaiger & Kemp, 1990):

$$DREE = \frac{\dot{V}O_2 \times \Delta_K H_{O_2}}{1000}$$

**with** *DREE* = daily resting energy expenditure in $\frac{kJ}{h \cdot g}$ ; $\dot{V}O_2$ = oxygen consumption in $\frac{ml\ O_2}{h \cdot g}$ ; $\Delta_K H_{O_2}$ = oxycaloric equivalent (here: 20.37 $J/ml$ ); divided by 1000 to convert DREE units from *J h$^{-1}$ g$^{-1}$* to *kJ h$^{-1}$ g$^{-1}$*.

Mean body mass (BM) was included for converting it into mass-specific metabolic rate (MR [$\frac{ml\ O_2}{h \cdot g}$]). Thereafter, the metabolic rate values were averaged over a 5-minute interval and then merged with corresponding temperature data ($T_{Skin}$, $T_{box}$ and $T_{ambient}$). It was not possible to derive a general MR threshold for torpor because the metabolic rate was individually variable, even within each site. Usually, torpor is also defined by a reduction of the body temperature to approximately 15 – 20°C (Tobler, 2011). Since a high diversity regarding $T_{skin}$ during torpor and a variety in torpor patterns in general (e.g. "hot" torpor with high $T_{skin}$) was observed, it was also not possible to derive and apply a $T_{skin}$ threshold for torpor. Thus, torpor was defined manually by checking the daily course of the metabolic rate for low values and periods with a very small amplitude of MR fluctuations. For calculating the MR reduction, first the MR of all periods in which the bats were not in a torpid state was defined as a day-resting metabolic rate ("DRMR"). Afterwards, the highest 20% of the DRMR values were removed, to exclude periods of arousals and other high activity phases, before calculating the MR reduction during a torpid state compared to the DRMR. The MR from all periods, except the torpor periods, could be defined as a resting metabolic rate, because bats exhibit the highest energy expenditure in their active phase during flight but were not able to fly during measurements in the metabolic chamber. Hence, all activities inside the metabolic chamber (e.g. changing the position) regardless of the time of the day were not very energy consuming and the MR was much closer to a resting state.

The bats' condition was calculated by dividing the body mass by the forearm length ($\frac{BM\,[g]}{FL\,[mm]}$). Although the BM is included in the physical condition, the body mass and the bats' condition were used separately for statistical analyses. The BM can be used to check for differences in energy that is expended to supply the specific body mass itself, whereas the condition indicates the bats' biological fitness and the amount of energy that can be provided.

## *Statistical analyses*

Data were combined, processed and analyzed using CRAN R (R Core Team, 2018) and the following packages: "RStudio" (R Studio Team, 2016), "lubridate" (Grolemund & Wickham, 2011), "car" (Weisberg & Fox, 2011), "ggplot2" (Wickham, 2009), "lattice" (Sarkar, 2008), "dplyr" (Wickham et al., 2020), "zoo" (Grothendieck & Zeileis, 2005), "plyr" (Wickham, 2011), "readxl" (Wickham & Bryan, 2017) and "data.table" (Dowle & Srinivasan, 2017).

First step in data analyses was testing for normal distribution of the data using the Shapiro Wilk test. Data were tested for differences between means for independent samples and for correlations between variables. When variables were normally distributed, the Welch two-sample t-test, one-way ANOVA (followed by a Tukey post hoc analysis) and two-way ANOVA (if checking for 2 Variables and >2 levels) were used. When variables were not normally distributed, the Kruskal-Wallis rank sum test (followed by a Dunn's test post hoc analysis, if data contained 2 Variables and >2 levels) was used. The Rayleigh test of uniformity was used for testing the circular distribution of the torpor entry times, since we converted the time of entry in decimal time (Jammalamadaka & SenGupta, 2001). The Pearson correlation test was applied for normally distributed data and the Spearman rank correlation test, if data were not normally distributed. The chosen test for each specific case was declared directly in the results. Data are presented as mean values ± standard deviation; *N* represents the number of individuals; *n* is the number of total measurement points included.

# Results

In total, the oxygen consumption and skin temperature of 39 individuals were measured (15 females, 24 males) under stable and buffered cave conditions (Andranolovy: $N = 16$) and under fluctuating environmental conditions in the other cave (Vintany: $N = 12$) and in a forest (Kirindy: $N = 11$). There was a significant difference among the three study sites regarding the bats condition (one-way ANOVA; $F_{2,39} = 4.34$, $P = 0.02$) and BM (one-way ANOVA; $F_{2,39} = 4.69$, $P = 0.01$). A Tukey post hoc analysis reported only a difference between Vintany and the other two sites (condition: $P = 0.01$; BM: $P = 0.03$), but none between Andranolovy and Kirindy (Table 2).

The environmental temperature and relative humidity of the three study sites varied markedly regarding the long-term measurements (Fig. 4 a-c). A significant difference among daily mean temperatures (Kruskal-Wallis-test; $X^2_{2,148} = 99.17$, $P < 0.01$) and daily relative humidity (one-way ANOVA; $F_{2,148} = 85.02$, $P < 0.01$), as well as daily fluctuations of both (Kruskal-Wallis-test; temperature: $X^2_{2,148} = 131.11$, $P < 0.01$; humidity: $X^2_{2,148} = 130.2$, $P < 0.01$) could be identified a for all sites. Andranolovy, the well-buffered cave with daily fluctuations less than 0.03°C and 1.4% RH (Table 1), was the hottest cave with a temperature and RH never below 29.4°C and 77%, respectively.

Table 1: Environmental temperature and relative humidity of each study site. Mean values and the standard deviation in °C, calculated out of daily measurement data from each study site. $T_{range}$ and $RH_{range}$ are described by the mean minimum and maximum values.

|  | *Andranolovy* (n = 47 days) | *Vintany* (n = 47 days) | *Kirindy* (n = 56 days) |
|---|---|---|---|
| $T_{mean}$ | 29.39 ± 0.04 | 23.05 ± 1.62 | 22.07 ± 1.44 |
| $RH_{mean}$ | 87.44 ± 8.04 | 75.02 ± 5.78 | 68.68 ± 7.65 |
| $T_{range}$ | 29.38 ± 0.04 – 29.41 ± 8.17 | 19.19 ± 2.05 – 26.83 ± 1.87 | 13.33 ± 2.63 – 33.98 ± 1.81 |
| $RH_{range}$ | 86.71 ± 7.86 – 88.16 ± 8.17 | 59.36 ± 9.17 – 87.11 ± 5.0 | 27.52 ± 7.43 – 95.20 ± 5.9 |
| $T_{fluc}$ | 0.03 ± 0.03 | 7.64 ± 1.97 | 20.65 ± 3.27 |
| $RH_{fluc}$ | 1.45 ± 0.68 | 27.75 ± 8.39 | 67.67 ± 8.44 |

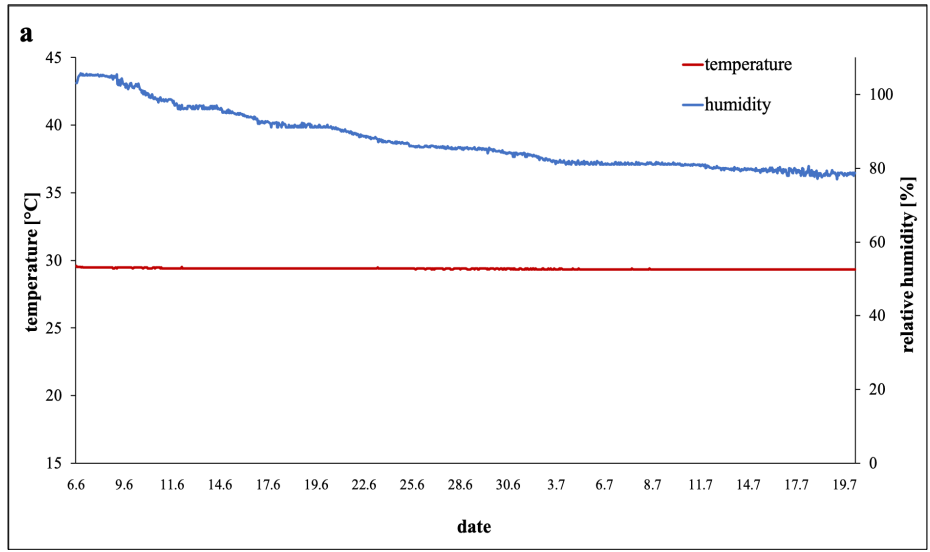

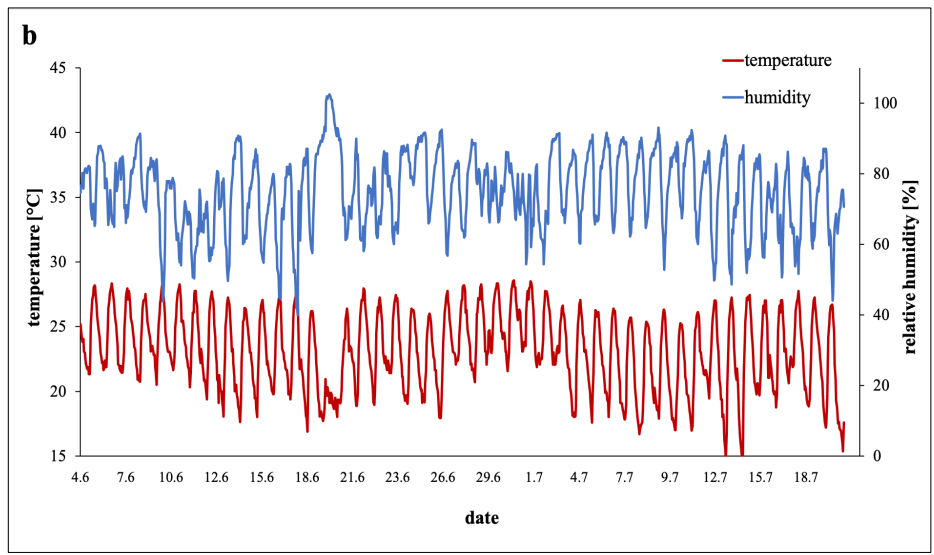

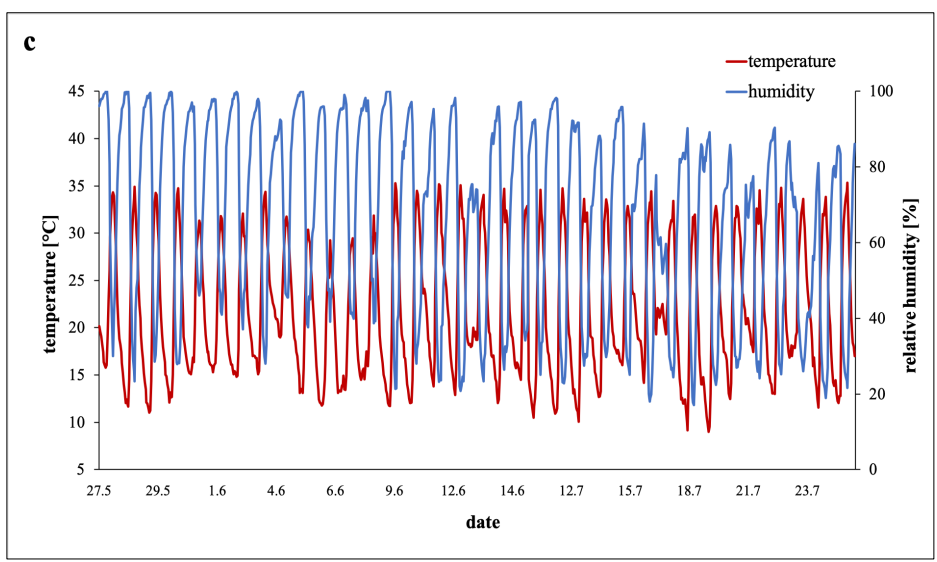

*Figure 4: Environmental temperature (red) and relative humidity (blue) long-term measurements in Andranolovy cave (**a**), Vintany cave (**b**) and Kirindy forest (**c**).*

Vintany cave exhibited a daily mean temperature of 23.05±1.6 °C and 75±5.8% RH daily mean (Table 1). It was more affected by ambient conditions given its cave structure and showed clear day and night fluctuations (Fig. 4 b). In the forest habitat Kirindy, temperature and RH were the lowest out of all three study sites ($T_{min}$: 13.3±2.6°C, $RH_{min}$: 27.5±7.4%) and displayed the highest variation in daily fluctuations, up to 28.4°C and 84.8% RH (Fig. 4 c).

Overall, 62% ($N$=39) of the bats entered a torpid state under varying ambient conditions and showed great differences regarding metabolic rate and oxygen consumption during torpor. Furthermore, associated skin temperatures, the duration, and different patterns of the torpor varied considerably. While all measured bats in Kirindy entered torpor ($N$ = 11), only 58.3% in Vintany ($N$ = 12) and 37.5% in Andranolovy ($N$ = 16) did. Bats reduced their metabolic rate during a torpid state by up to 96% of their DRMR. The metabolic reduction differed significantly among all sites (one-way ANOVA; $F_{2,24}$ = 46.6, $P < 0.01$), but a Tukey post Hoc analysis uncovered no significance between the direct comparison of Vintany and Kirindy ($P = 0.3$; Tab. 2).

Bats roosting in the forest showed the lowest TMR with a minimal rate of 0.05±0.03 ml $O_2$ per hour and gram body weight, but also the highest metabolic rate in general with a maximum metabolic rate of 10.08±2.23 ml $O_2$ $h^{-1}$ $g^{-1}$ (Fig. 5). There was a significant difference among all study sites regarding the TMR (Kruskal-Wallis-test; $X^2_{2,223}$ = 8.59, $P = 0.01$) and also the DRMR (Kruskal-Wallis-test; $X^2_{2,2682}$ = 131.76, $P < 0.01$). A direct pairwise comparison of the TMR using the "Dunn's test" revealed a slight difference of means, yet only moderate significant between the well buffered cave and the forest habitat ($P = 0.03$). The direct comparison of the TMR between the bats from the two caves did show similarities ($P = 0.9$), whereas the bats' TMR in the more open cave and the forest habitat expressed differences, but no significance ($P = 0.07$). In contrast to these results, the Dunn's test comparing the DRMR showed highly significant differences between each site, respectively ($P = <0.01$). The mean DRMR was the lowest in Andranolovy with 2.66±1.05 ml $O_2$ $h^{-1}$ $g^{-1}$, compared to the other sites (Fig. 5).

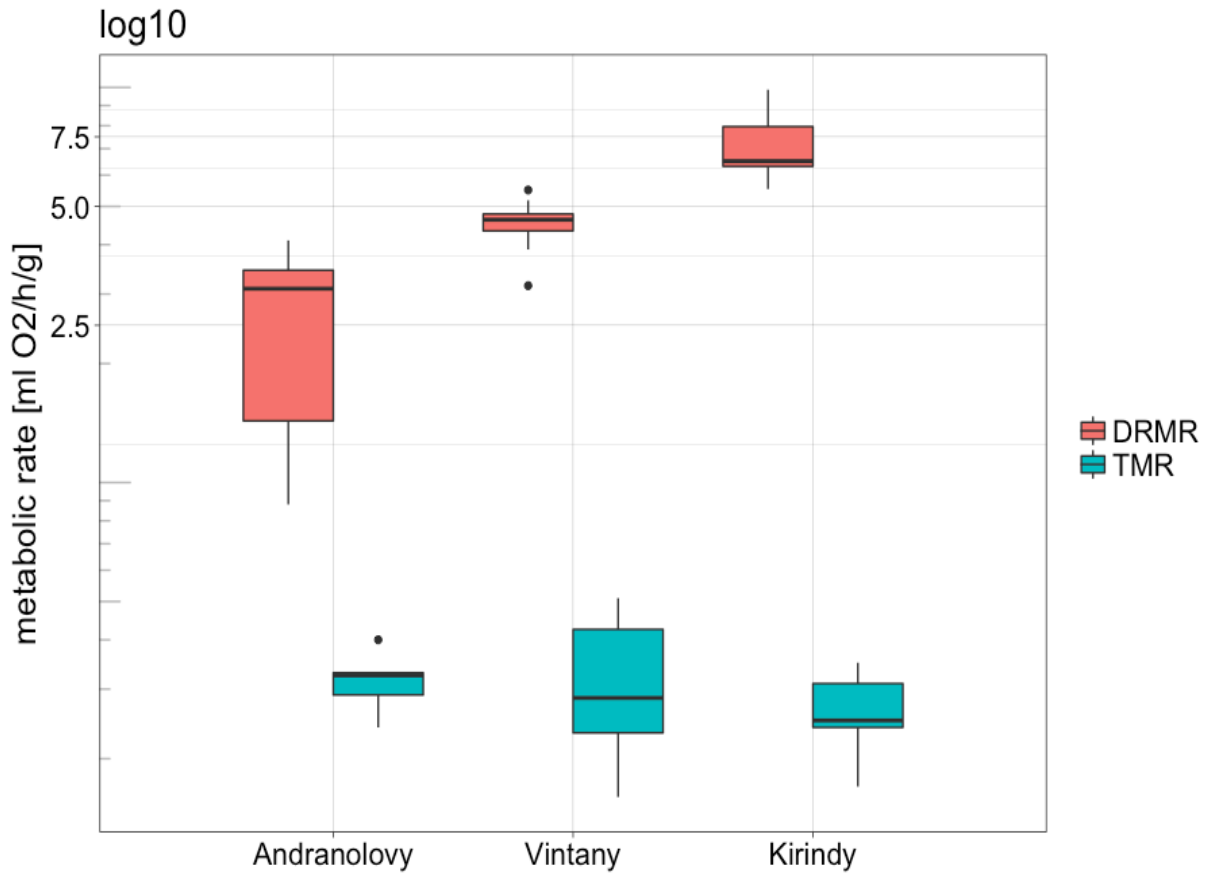

*Figure 5: Metabolic rate [ml $O_2$ $h^{-1}g^{-1}$] for "DRMR" (daily resting metabolic rate) and "TMR" (torpid metabolic rate), grouped for each site. The Y-axis was "$log_{10}$"- transformed to account for the wide range between "DRMR" and "TMR".*

Further tests were taken to examine potential effects of the bat's sex, BM, condition and their interactions on the metabolic rate. Only the bats' condition and BM as main effects were statistically significant concerning the DRMR (two-way ANOVA[condition]: $F_{1,39}$ = 11.35, $P < 0.01$; two-way ANOVA[BM]: $F_{1,39}$ = 8.13, $P < 0.01$), whereas none of the effects or interactions were significant for the TMR.

*Table 2: Comparison of metabolic rate, skin temperature and the bat's body characteristics between the study sites (mean values ± standard deviation were calculated out of daily mean values from each individual). Units are shown on the left or directly in rows. The MR reduction was calculated by removing the highest 20% of the DRMR (to exclude periods of arousals and other extreme high activity phases) before computing the share of the TMR.*

| | | ***Andranolovy*** $N_{(DRMR)} = 16$ $N_{(TMR)} = 6$ | ***Vintany*** $N_{(DRMR)} = 12$ $N_{(TMR)} = 7$ | ***Kirindy*** $N_{(DRMR)} = 11$ $N_{(TMR)} = 11$ |
|---|---|---|---|---|
| *[ml $O_2$ $h^{-1}$ $g^{-1}$]* | DRMR | 2.66 ±1.05 | 4.53 ±0.6 | 7.21 ±1.38 |
| | TMR | 0.32 ±0.05 | 0.31 ±0.14 | 0.27 ±0.06 |
| | MR min | 1.10 ±0.97 | 0.17 ±0.14 | 0.05 ±0.03 |
| | MR max | 5.46 ±1.19 | 7.82 ±0.74 | 10.08 ±2.23 |
| | MR reduction [%] | 70.98 ±9.33 | 91.78 ±4.02 | 95.13 ±1.65 |
| *[°C]* | $T_{skin}$ normo | 36.01 ±1.89 | 31.76 ±2.55 | 29.93 ±4.15 |
| | $T_{skin}$ torpid | 34.21 ±4.29 | 24.80 ±2.67 | 24.19 ±6.54 |
| | $T_{skin}$ min | 32.95 ±3.13 | 24.17 ±3.86 | 16.81 ±8.28 |
| | $T_{skin}$ max | 38.48 ±0.88 | 38.19 ±1.04 | 36.70 ±0.99 |
| | $T_{skin}$ range | 5.54 ±2.94 | 14.02 ±3.27 | 19.89 ±8.29 |
| | BM [g] | 10.42 ±1.14 | 10.50 ±1.19 | 9.26 ±0.8 |
| | FL [mm] | 47.98 ±2.29 | 50.62 ±1.88 | 46.93 ±1.1 |
| | condition [g/mm] | 0.22 ±0.02 | 0.20 ±0.02 | 0.21 ±0.02 |

Even though there was neither a significant difference in the time of torpor entry between the different habitat forms of cave and forest (Kruskal-Wallis-test; $X^2_{2,24} = 2.93$, $P = 0.09$) nor between each site (Kruskal-Wallis-test; $X^2_{2,24} = 4.36$, $P = 0.12$), it was observable that bats roosting in the forest entered torpor slightly earlier than the ones roosting in caves (Fig.6). The Rayleigh-test for circular statistics revealed, that the entry time is significantly orientated in a specific direction ($\bar{R} = 0.49$, $P < 0.01$), since the bats entered torpor predominantly in the early mornings. The mean resultant length ($\bar{R}$) is a measure of spread around the circle, with values between 0 (completely spread) and 1 (concentrated on one point). Further examinations displayed, that individuals in Kirindy forest showed the widest range of torpor entry time (~ 3.5 hours; Rayleigh test: $\bar{R} = 0.4$, $P = 0.17$), whereas the torpor entry times of bats in Andranolovy (Rayleigh test; $\bar{R} = 7.4$, $P = 0.03$) and Vintany (Rayleigh test; $\bar{R} = 0.64$, $P = 0.054$) were more restricted (~2.5 hours). Yet, only the results from bats in Andranolovy were significant, due to statistical outliers in Kirindy and Vintany.

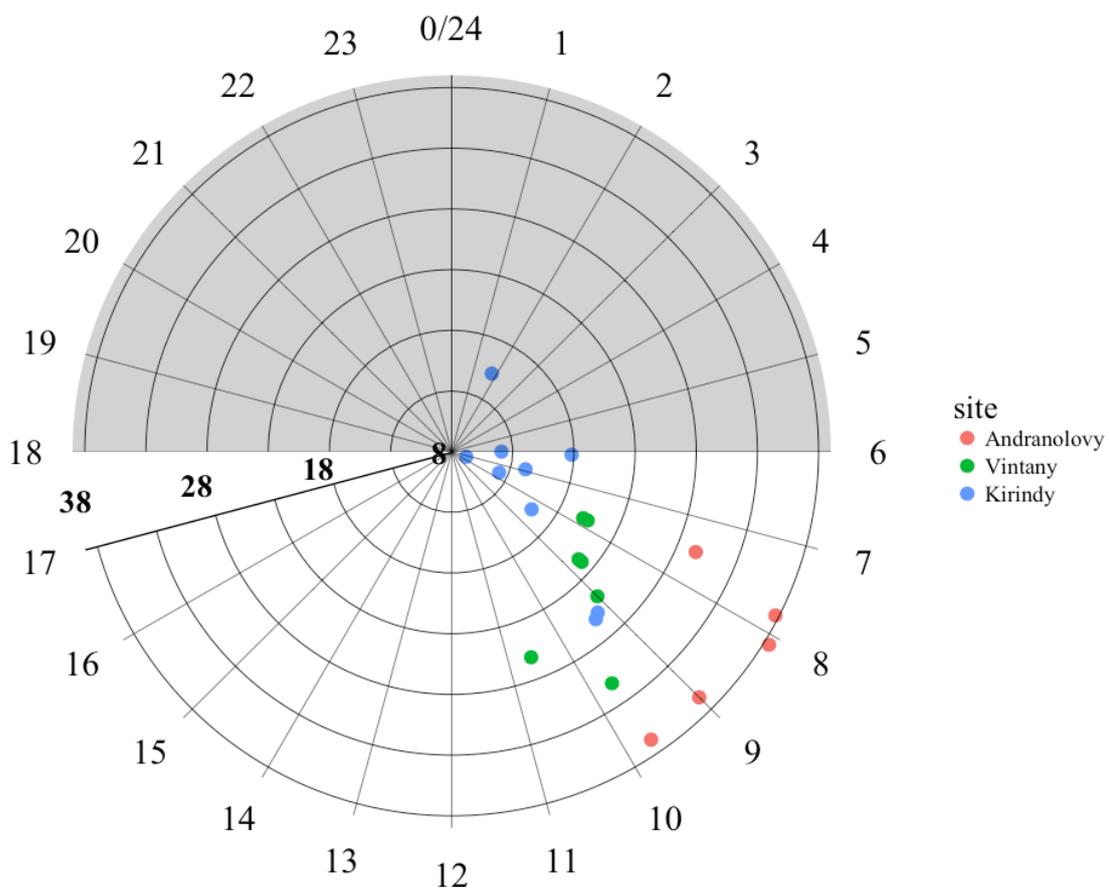

Figure 6: Circular distribution of the skin temperature (radial distance) and the time (degrees) of torpor entry. Labels around the circle denote the time of the day (format HH) and shaded areas indicate the dark/night periods. The temperature axis on the left is illustrated in 5°C steps from 8 – 38°C.

Moreover, skin temperature at the time of torpor entry differed significantly for each site (one-way ANOVA; $F_{2,24}$ = 31.19, $P < 0.01$). In the well buffered cave Andranolovy the skin temperature of the individuals that entered torpor was mostly between 35 and 38°C ($N = 6$), whereas the skin temperature of the forest dwelling individuals in Kirindy was usually around 15°C ($N = 11$). Skin temperature at torpor entry of the bats roosting in the sinkhole with more fluctuating ambient conditions (Vintany) was in the range between 19 – 30°C ($N = 7$) (Fig.6).

The course of $T_{skin}$ and MR revealed a day and night pattern in which all individuals, regardless of entering a torpid state or not, reduced both ($T_{skin}$ and MR) during the daytime (Fig. 7). The normothermic skin temperature was the highest in Andranolovy under stable environmental conditions (Table 2). This microhabitat differed highly from the other two habitats regarding $T_{skin}$ (one-way ANOVA; $F_{2,39}$ = 16.15, $P < 0.01$), whereas Vintany cave and Kirindy forest did not display differences between each other (Tukey post-hoc analysis [Vintany-Kirindy]: $P = 0.3$). The pattern of $T_{skin}$ of every individual showed disparities within each site (Fig. 7a). While bats in Andranolovy roosted under constantly high ambient temperatures and therefore did not have a high variance in $T_{skin}$ (Table 2), the range of skin temperature of the bats in Kirindy was immense (up to 20°C during a one-day measurement), given the fluctuating environmental conditions (Fig. 7b). Just like the normothermic skin temperature, the differences in the torpid $T_{skin}$ were only significant between Andranolovy and the other two microhabitats (one-way ANOVA; $F_{2,24}$ = 9.14, $P < 0.01$), however, bats from Vintany and Kirindy had almost the same skin temperature during torpor (Tukey post-hoc analysis [Vintany-Kirindy]: $P = 0.96$) (Table 2).

Furthermore, it was also tested whether there are any differences in $T_{skin}$ with regard to the bats sex, as well as for a possible correlation between $T_{skin}$ and the bat's condition. Even though the males' skin temperature was on average about two degrees [°C] higher than that of the females, this inequality was neither significant during torpid periods (Welch two-sample t-test; $t_{(1,24)}$ = -0.9, $P = 0.34$), nor during normothermic periods (Welch two-sample t-test; $t_{(1,39)}$ = -1.8, $P = 0.08$). Moreover, there were significant positive correlations between the bat's condition and their skin temperature during both periods (Pearson correlation; [normothermic] $r_{22} = 0.5$, $P < 0.01$; [torpor] $r_{22} = 0.46$, $P = 0.02$), but correlation coefficients were moderate.

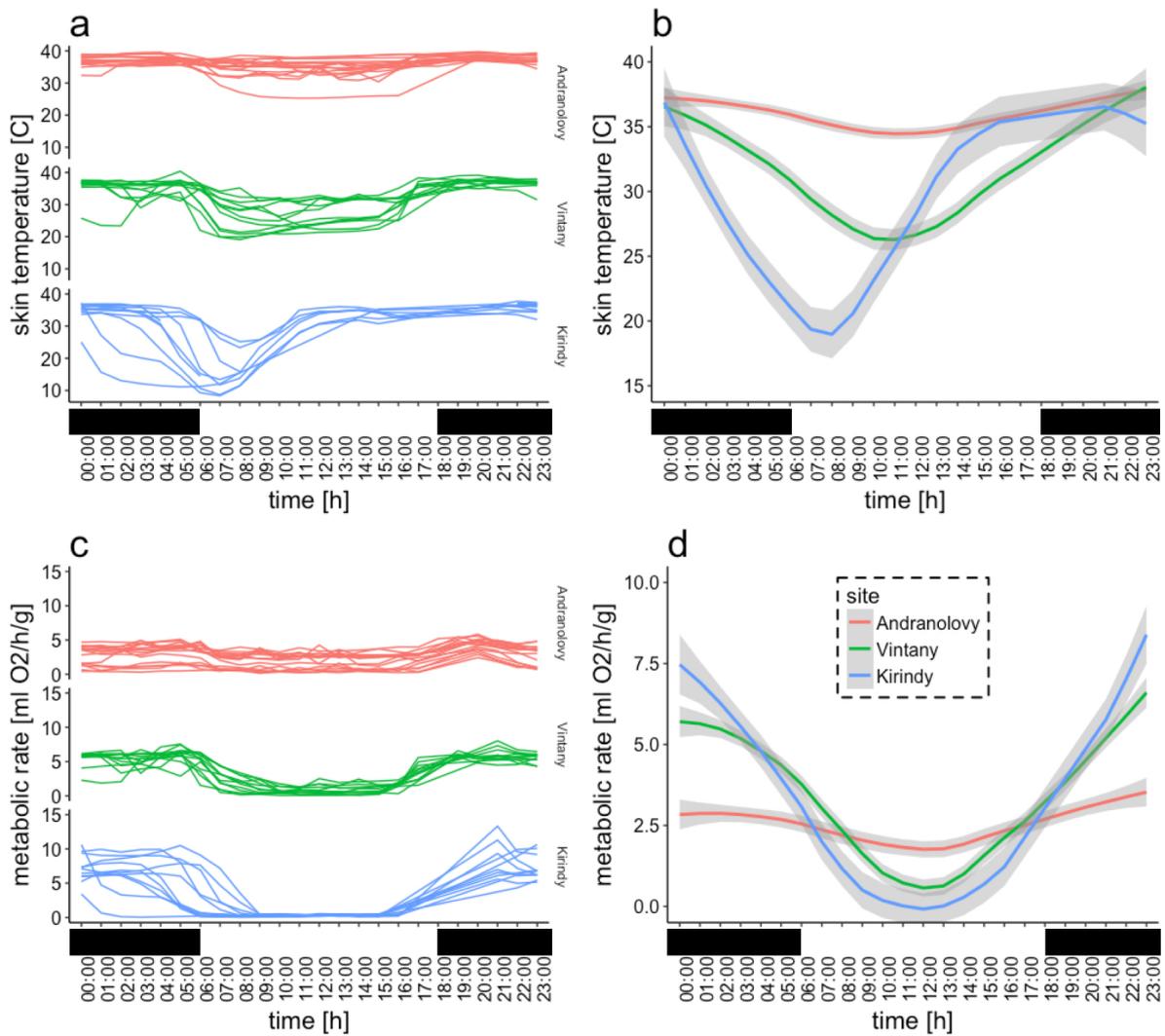

*Figure 7: Skin temperature layered from all individuals (a) and mean skin temperature with standard deviation as shaded area around the lines (b) during the course of a day and separated by sites (Andranolovy = red, Vintany = green, Kirindy = blue). The same visualization applies for the metabolic rate from all individuals (c) and for mean values per site (d). The bars below the X-axis indicate the differences between day and periods (day = white; night = black).*

The similarities in the individual patterns of $T_{skin}$ and MR (Fig. 7a and c) indicate a positive correlation between both, which was highly significant during a one-day measurement (Spearman correlation; $r_s = 0.45$, $n = 786$, $P < .001$). However, an in-depth examination of only the torpid periods revealed no significant correlation between $T_{skin}$ and TMR (Pearson correlation; $r_{22} = 0.07$, $P = 0.3$). Although Vintany is, like Andranolovy, a cave habitat, the bats' daily pattern of MR is much more similar to that from the bats in the forest habitat Kirindy (Fig. 7d).

The comparison of the duration of torpor of all individuals across all sites exemplifies a significant difference (one-way ANOVA; $F_{2,24}$ = 6.04, $P$ < 0.01). Individuals from Andranolovy cave exhibited the shortest torpor duration, yet a high degree of variance (4.8±3.1 hours). The mean torpor duration was the highest in Kirindy forest, were individuals stayed in a torpid state for 9.0±2.4 hours during a day-long measurement (Fig. 8). A Tukey post hoc analysis revealed that only the sites Andranolovy and Kirindy differed significantly at $P$ < 0.01. Torpor duration of the individuals from Vintany was not significantly different from the other two groups, lying in the middle of both (6.4±1.9 hours).

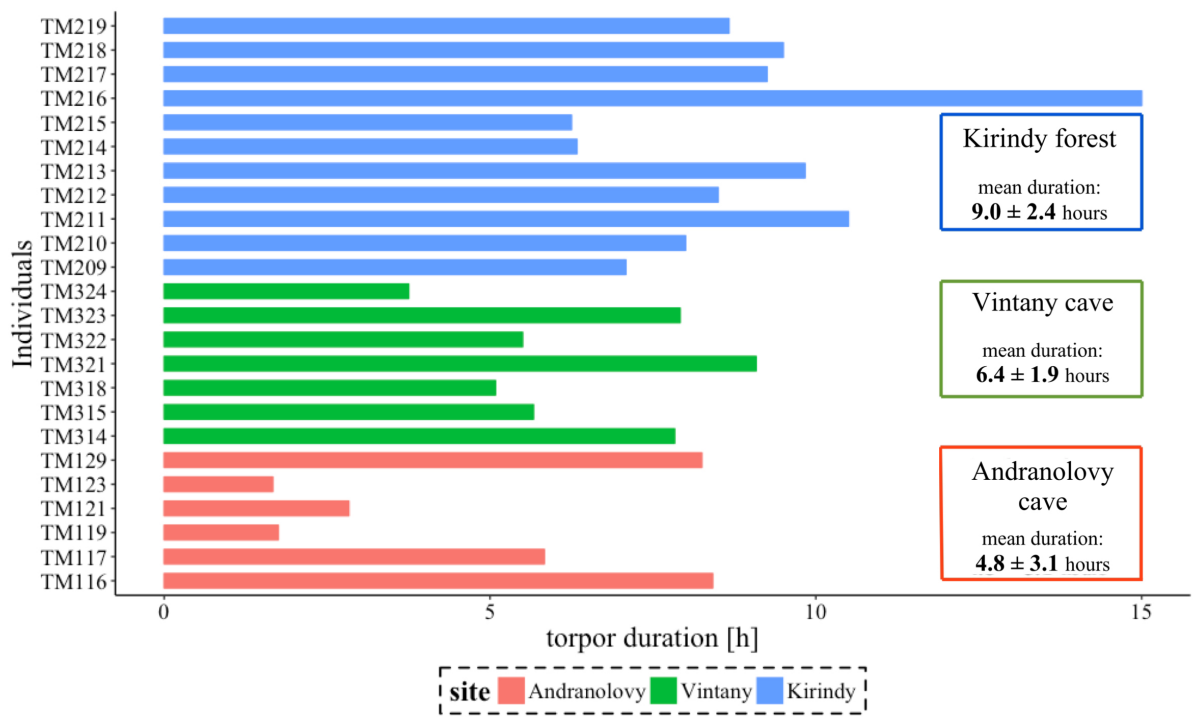

*Figure 8: Torpor duration (in hours) for each individual that entered a torpid state. Y-axis shoes individually ID's grouped for associated sites. Text blocks on the right visualize the mean torpor duration and standard deviation for each group.*

Tests to externalize differences in sex and a potential interaction between the roosting site and the sex regarding the torpor duration have also been conducted. None of these revealed a significant effect on the torpor duration (two-way ANOVA [sex]: $F_{1,24}$ = 2.40, $P$ = 0.14; two-way ANOVA [sex:sites]: $F_{2,24}$ = 1.57, $P$ = 0.24) .

Results of the Pearson correlation indicated no significant negative relations between the bats' body mass (BM) and torpor duration (Pearson correlation; $r_{22}$ = -0.05, $P$ = 0.8), or their condition and torpor duration (Pearson correlation; $r_{22}$ = -0.2, $P$ = 0.36), respectively.

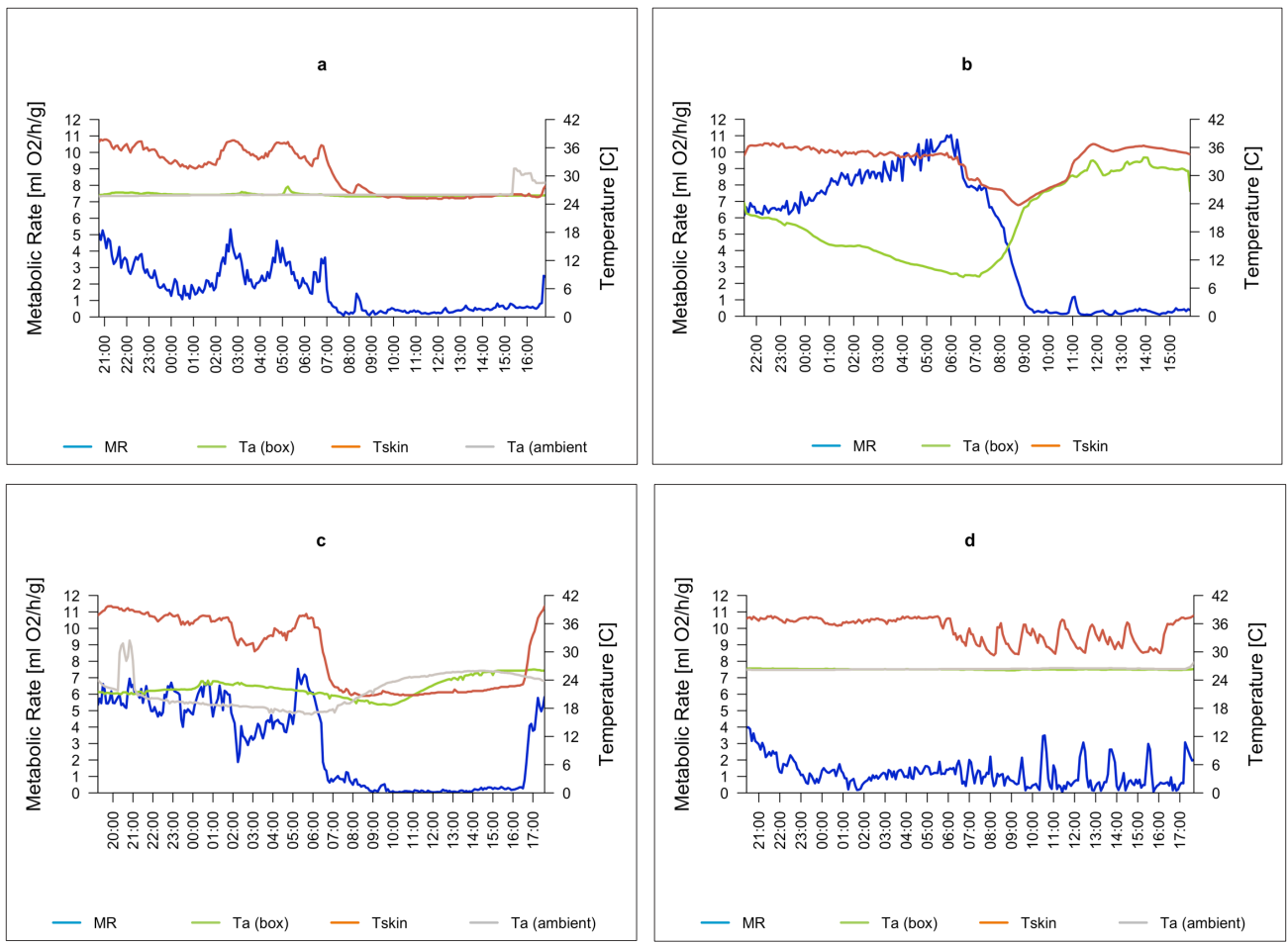

*Figure 9: MR, T$_{skin}$, T$_{box}$ (temperature inside the measurement box) and T$_{ambient}$ during the course of a one-day measurement from different individuals, showing variations in the torpor patterns. Paradigmatic selection for daily torpor in Andranolovy (**a**) and Vintany (**c**), as well as for "hot" torpor in Kirindy (**b**) and short torpor bouts in Andranolovy (**d**).*

Not only did the torpor duration or the hypometabolic range differ between each site and individual, but also did the torpor patterns show variations (Fig.9). Most of the torpid individuals ($N = 24$) showed a daily torpor pattern ($n = 16$), yet other patterns occurred as well, like a "hot torpor" pattern ($n = 6$) and unusually short torpor bouts ($n = 2$). The torpor patterns were closely linked to the environmental and ambient conditions, since the varying patterns were often related to the different types of microhabitats.

Patterns of the daily torpor mostly displayed a slowly entering hypometabolic state and an associated drop in skin temperature about 10 °C, until it reached ambient temperature, thenceforth $T_{skin}$ was stable on ambient level (Fig. 9a). In Vintany, were bats were more exposed to temperature fluctuations, $T_{skin}$ dropped to 18°C during daily torpor and remained constant even below $T_{ambient}$ (Fig. 9c). Most of these individuals started entering torpor in the early morning hours (~02:00) but increased MR and $T_{skin}$ again slowly, until they rapidly entered torpor completely around 07:00 in the morning.

The uncommon "hot" torpor occurred mostly in Kirindy, given the high ambient temperatures during the daytime. As in this example from an individual from Kirindy, the bat increased MR excessively against decreasing $T_{box}$ (for measurements in Kirindy: $T_{box} = T_{ambient}$) during the active phase at night, for keeping $T_{skin}$ stable, until it could not hold MR and $T_{skin}$ on a high level and decreased both constantly until entering torpor around 09:00 h (Fig. 9b). Since the ambient temperature in the forest increased already in the morning, $T_{skin}$ and $T_{box}$ matched at some point and $T_{skin}$ increased again together with $T_{box}$ (Fig. 9b; 08:30 h). This led to a "hot" torpor with unexpected high $T_{skin}$, while TMR was about 0.06 ml $O_2$ $h^{-1}$ $g^{-1}$.

Short torpor bouts were only expressed by bats from the well buffered cave Andranolovy. Periods of a torpid state only lasted for about 10 to 30 minutes and were always followed by actions of arousals accompanied by high oxygen consumption (Fig. 7d). $T_{skin}$ followed the same pattern as the TMR; after it slowly decreased, there was always a rapid and short increase ($T_{skin}$ range of ~ 8°C), even though ambient temperatures were constantly stable around 26°C. The lowest torpid $T_{skin}$ was 7.9°C with a TMR of 0.67 ml $O2$ $h^{-1}$ $g^{-1}$, recorded for an individual from Kirindy forest during the early morning hours. The highest torpid $T_{skin}$ was 39.4°C with a TMR of 0.15 ml $O2$ $h^{-1}$ $g^{-1}$, recorded from a bat from Andranolovy cave that expressed (besides one other bat from the same cave) a combination of a "hot" torpor pattern with short torpor bouts.

For comparing the daily resting energy expenditure (in kJ $^{-h\ -g}$) among the study sites, the DREE values (in 5–minute intervals) from each individual were added up for each study site. To account for different numbers of individuals per site the sum was divided by *N*, resulting in individuals' mean DREE per site (Table 2). Data was partitioned into DREE at night and during the day, to check for differences during the active and inactive phase.

The general DREE was the highest in bats from Vintany, whereas bats from Andranolovy and Kirindy had almost the same energy expenditure. However, the sites did not differ significantly (Kruskal-Wallis-test; $X^2_{2,36}$ = 3.37, $P < 0.18$). During the day, the inactive phase, bats in Kirindy spent the least energy on average (0.15 ± 0.003; Table 3). Bats in Vintany spent more than twice as much energy and individuals in Andranolovy showed the highest DREE$_{day}$ (0.46 ± 0.002). While there was a significant difference among all sites (Kruskal-Wallis-test; $X^2_{2,36}$ = 12.44, $P < 0.01$), a post hoc test revealed, that only Andranolovy and Kirindy differed significantly (Dunn's test: *P = 0.01*) The DREE$_{night}$ was reversed, since bats in Andranolovy spent the leas energy during the active phase and DREE$_{night}$ differed significantly from the other two sites (one-way ANOVA; $F_{2,36}$ = 14.33, $P < 0.01$). The daily resting energy expenditure in bats from Vintany was slightly higher than in bats from Kirindy, yet there was no significant difference between both sites (Tukey post-hoc analysis: *P = 0.9*) (Table 3).

*Table 3: Daily resting energy expenditure (in kJ $h^{-1}$ $g^{-1}$) of each study site. DREE$_{day}$ (06:00 – 18:00) and DREE$_{night}$ (18:00 – 06:00) were classified by approximate sunrise and sunset.*

|  | ***DREE*** | ***DREE**$_{day}$* | ***DREE**$_{night}$* |
|---|---|---|---|
| ***Andranolovy*** *(N=16)* | 1.05 ± 0.003 | 0.46 ± 0.002 | 0.60 ± 0.003 |
| ***Vintany*** *(N=12)* | 1.41 ± 0.004 | 0.36 ± 0.002 | 1.06 ± 0.002 |
| ***Kirindy*** *(N=11)* | 1.06 ± 0.006 | 0.15 ± 0.003 | 1.01 ± 0.005 |

# Discussion

We compared different populations of the bat *Triaenops menamena* and found, that rather the gradient of fluctuation in environmental conditions was decisive for different metabolic rates, torpor patterns, torpor duration or reduction rates than the type of habitat itself. Indeed, Andranolovy and Vintany are both cave habitats, yet they show more dissimilarities within each other than the Vintany cave compared to the forest habitat. The metabolic reduction rate, the TMR and skin temperature for DRMR and TMR from bats in Kirindy forest and Vintany cave resembled each other more, than the two sites in comparison to bats from the well-buffered cave Andranolovy. These two microhabitats are much more exposed to ambient conditions and are subjects to daily fluctuations, which explains the resemblances of the results. The rapid change in temperature and RH are highly influential for the bats' physiological stress and their response to it, by entering a torpid state to escape unfavorable conditions (Kobbe et al., 2014). Nevertheless, the immense differences between Andranolovy and Vintany were surprising, since the two caves are less than 300 meters apart and individuals from the two populations could theoretically roost in both caves.

For this thesis, it was presumed that the bats roosting in more fluctuating microhabitats, like Kirindy forest, generally expend less energy, than the bats roosting in more stable conditions, e.g. inside caves, due to increased occurrence of torpor and a greater reduction in the metabolic rate. In general, caves can provide thermally stable and humid environmental conditions and protection against unpleasant weather transformations (Avila-Flores & Medellin, 2004). In tropical regions, many bat species are roosting in hot caves throughout the year, which are either heated by convection of hot air or by the bats themselves (Rodríguez-Durán & Soto-Centeno, 2003; Furey & Racey, 2016). Caves with less ambient temperature variability show significantly higher bat species richness (Cardiff, 2006) and caves with multi-species assemblages are often inhabiting more individuals than caves with monospecific colonies, because of interspecific variations in foraging patterns (Rodríguez-Durán, 1998). Hence, tropical caves are often populated by great colonies, aggregating in large numbers, which can lead to thermoregulatory advantages (e.g. trapping body heat by clustering or a decreased evaporative water loss) and reduce predation risk due to the high densities of bats. However, great colonies may also bring disadvantages, like an increased disease transmission and higher incidence of parasites, as well as a potential higher interspecific foraging competition (Furey & Racey, 2016).

Andranolovy cave inhabits such a multi-species assemblage and is the hottest and most buffered roost out of all three study sites. With a daily mean temperature of ~30°C and a mean relative humidity of 87%, it most probably complies with the conditions of the bats' thermoneutral zone. Within the bats' TNZ, normothermic thermoregulation at minimal costs would be possible, while still being able to perform normal functioning physiological and behavioral activities, which is beneficial at a resting state (Turbill, 2006). The reduced costs arise from the constantly high temperature and water saturation in the air, since bats do not have to actively thermoregulate and spend energy for maintaining a high body temperature or reducing their evaporative water loss (EWL) (Baudinette et al., 2000). This is reflected in the results, as bats in Andranolovy expressed the highest skin temperature during normothermic and torpid states (never below 32.95±3.13°C on average) and expended the least energy during the active phase (0.60±.003 kJ $h^{-1}$ $g^{-1}$ on average), since they are very likely dwelling within their TNZ. Even though the energy the bats have to expend for obtaining normothermia was the lowest in Andranolovy cave, torpor is not only used as a response to acute energy shortage but also to regularly balance the overall energy budget (Geiser, 2003). Consequently, not only did the bats, that roost in the more fluctuating microhabitats enter torpor and reduced their metabolic rate remarkably, but so did the bats in Andranolovy.

Vintany as well is a cave habitat, but much more affected by fluctuating ambient conditions with day-and-night patterns, because of its open structure as a sink hole. Therefore, more bats entered torpor to save energy as a response to the more fluctuating conditions (Lovegrove et al., 2014) compared to Andranolovy cave. Bats in Vintany have to cope with periods of unfavorable ambient conditions during the day and, if they do not become torpid, have to spend more energy for keeping the body temperature on a steady high level. Bats roosting in Kirindy forest displayed the highest metabolic rate, on average 7.21 ±1.38 ml $O_2$ $h^{-1}g^{-1}$, for staying normothermic in this microhabitat, where daily ranges of up to 28.4°C and 84.8% RH occur. The high thermoregulative costs and fluctuating ambient conditions during the day caused 100% of the bats in Kirindy entering torpor, resulting in a lower DREE (0.15±0.003 kJ $h^{-1}$ $g^{-1}$ on average) compared to bats in Vintany (0.36±0.002 kJ $h^{-1}$ $g^{-1}$ on average), where only 58,3% used torpor to escape unfavorable ambient conditions. The metabolic rate reduction during torpor was the highest in Kirindy forest, where bats reduced their DRMR by ~95% on average. Bats in Vintany cave showed a similar reduction rate, while entering torpor. Yet, the lowest TMR on average (0.27 ±0.06 ml $O_2$ $h^{-1}g^{-1}$) and the lowest metabolic rate value in general (0.018 ml $O_2$ $h^{-1}g^{-1}$) were expressed in the forest. Compared to frugivorous and nectivorous bats, the metabolic rate of insectivorous bats, like *T. menamena*, is generally lower (Rodríguez-Durán, 1995), most likely due to the availability of flying insects year-round, even in tropical regions (Geiser, 2004), but it might also be influenced by the diet itself, since nectar and fruits provide a high sugar composition (Wolff, 2006).

There was a noticeable gradation in the bats' DRMR and its reduction to TMR among all sites: the greater the environmental conditions were fluctuating, the higher was the general DRMR and the rate of its reduction, but the lower was the TMR. The divergences in the environmental conditions between the well-buffered cave and the more fluctuating sites is also reflected in the results of the bats' skin temperature. That is, bats' skin temperatures were similar in Vintany cave and Kirindy forest and only differed in comparison to Andranolovy, regardless of their physiological state. The overall skin temperature was the highest in bats from Andranolovy, due to the high and stable cave temperature. Yet, the trapped body heat from clustered bats in this large colony might also be influential (Roverud & Chappell, 1991). Additionally, the bats' body mass and their condition (relation between BM and FL) had significant effects on their DRMR, i.e. the "fitter" the bats are, the more energy they can spend on maintaining normothermic, before entering torpor as an emergency response when saving energy becomes vital for survival. Previous studies on the free-ranging sugar glider *Petaurus breviceps* also revealed the ability to maintain normothermic for a longer time when in good condition (Körtner & Geiser, 2000). Furthermore, the bats' condition was positively correlated with their skin temperature, i.e. the better the bats' condition, the higher their body temperature.

Not only was the occurrence of daily torpor and the accompanying reduction of metabolic rate the highest in the least buffered microhabitat Kirindy, but also the duration of torpid states. Torpor duration varied significantly between all study sites and increased with greater ambient fluctuations along a gradient. Bats from the well-buffered cave in Andranolovy expressed the shortest torpor duration, whereas torpor lasted longer in the moderately fluctuating cave Vintany and was exceeded by bats from the forest in Kirindy. Other studies, like the ones about the tropical northern blossom-bat *Macroglossus minimus* revealed a temperature dependency of daily torpor, i.e. torpor lasts longer when ambient temperatures are decreasing (Bartels et al., 1998; Geiser 1994). Furthermore, torpor duration is also linked to the amount of energy expenditure spent during the torpid state. When ambient temperature is below the critical temperature of the TNZ, energy for thermoregulation and heat production must be dispensed, which results in shorter torpor bouts in general (Geiser & Kenagy, 1988). Even though the temperature is constantly high in Andranolovy cave, the inhabiting bats expressed the highest TMR and spent the most energy during the inactive phase (0.46±0.002 kJ h$^{-1}$ g$^{-1}$ on average) of all three microhabitats, which might be the reason for the comparably short torpor duration, yet high individual variance. Since the body temperature of bats in Vintany cave and Kirindy forest was decreasing with declining ambient temperatures, especially in the mornings, torpid states lasted longer in these microhabitats. Bats in Kirindy forest expressed the longest torpor duration as well as the lowest TMR and simultaneously faced the greatest environmental fluctuations and lowest temperature during the course of a day compared to the other study sites. Although, temperatures in the forest might be favorable at noon, possibly even allowing normothermia, the bats remained torpid. In this regard, a study about *Nyctophilus geoffroyi* – a bat species from Australia's tropical regions with nearly the same size and body mass like *T. menamena* – revealed a resting metabolic rate three times higher than the BMR when being normothermic at ambient temperatures of ~20°C. Hence, entering torpor provided substantial energy savings, even though roost temperatures were moderate in the late afternoon. (Geiser & Brigham, 2000). In tropical dry regions, the increased costs of evaporative water loss are also crucial for surviving. Thus, the comparatively long torpor bouts are accompanied by a reduction in EWL and substantial energy savings (Speakman & Thomas, 2003), which may outweigh the high costs of thermoregulation at night, when ambient temperature drops to 8°C, but the bats defend normothermia.

The bats' roost choice in Kirindy forest was also beneficial regarding the possibility of passive rewarming during the day after being torpid. Bats roosting in comparably exposed locations can save on energy for rewarming to a normothermic state while arousing from torpor by using the increasing ambient temperature for passive rewarming (Turbill et al., 2003). However, roost sites that allow passive rewarming, may often not provide thermal buffering in the afternoon or at night, when ambient temperatures are decreasing to a minimum. In some cases, the bats' skin temperature was not actively regulated during the torpid state, which led to daily T$_{skin}$ ranges of 19.89 ±8.29°C in Kirindy forest, since it followed ambient temperature fluctuations. Yet, bats roosting in the forest often might have no options of alternative higher-buffered roosts. Tree holes probably representing the highest insulation possible. Alternatives to escape unfavorable ambient conditions may be migration or performing hibernation, but the food availability is high and consistent year-round which allows a continuous activity (Geiser, 2004). Besides the longer torpor duration in the fluctuating habitats, a more synchronized entry time of torpor was also hypothesized. Sunrise is a strong indication for torpor entry in nocturnal heterotherms, as shown in the study about *Nyctophilus geoffroyi* (Turbill et al., 2003). The effect of dark-light cycles (photoperiodism) usually acts as the main *zeitgeber* for arousals and activity patterns (Francisand & Coleman, 1990). However, ambient temperature is also an effective determinant for the circadian rhythm and its effect gets even stronger in the absence of light stimuli (e.g. in caves) (Körtner & Geiser, 2000). Therefore, entering torpor in the early morning, when ambient temperature is normally lowest, appears to be most common for free-ranging bats (Audet & Fenton, 1988). Consequently, it was not only presumed that torpor and activity patterns of *T. menamena* are synchronized but, in particular, that bats roosting in the forest show a more synchronized time of torpor entry than bats from the caves, since they are more exposed to the effect of photoperiodism and ambient temperature.

Although bats in Kirindy forest tended to enter torpor slightly earlier at a much lower skin temperature, there was no significant difference in the time range of the individuals' torpor entry among all habitats. The *zeitgeber* for the earlier entry time for torpor in bats roosting in the forest was caused by the sunrise and the ambient temperature, which is coldest in the early mornings and not buffered by any habitat structures like in the sinkhole or the well-buffered cave. That is, bats in Kirindy were forced to enter torpor during periods of low ambient temperature but they also remained torpid during periods of low RH in the late afternoons, when temperature was rising. In this context, timing of torpor bouts coincided with minimum ambient temperature and low RH was also shown for the Gould's long-eared bat *Nyctophilus gouldi*, a tree-roosting bat inhabiting the (sub-)tropical regions of Australia (Turbill, 2006). The torpor entry time of most bats in the more exposed roosts was still relatively late (06:00 – 10:00 h) which contradicted the physiological expectations regarding the Arrhenius effect (Lovegrove, et al., 2014), considering that the coldest periods of the day were around 04:00 – 05:00 h in the morning.

However, not only the time of entry, but also the time of exiting a torpid state is important for the efficiency of daily torpor. Arousals triggered by sunset in the early evenings, when insect abundance and nightly temperature is highest, yields maximal foraging possibilities (Hickey & Fenton, 1996). Thus, Food availability and dark-light cycles (photoperiodism) and its linkage to ambient temperatures act as *zeitgebers* for the circadian rhythm for activity in animals, especially heterotherms (Francisand & Coleman, 1990). Yet, there is no effect on the synchronization of torpor entry from each individual roosting in the same microhabitat. Moreover, the larger sample size for torpid bats in Kirindy might have biased the results of the time range, since there were almost twice as many bats entering torpor than in the other two microhabitats, which might have led to a higher individually variety of entry times in the forest. Nevertheless, the hypothesis of the more synchronized time of torpor entry could not be confirmed by this study.

Roost choice is important for minimizing energy expenditure (whether bats are in a torpid or normothermic state) but roost preferences are not only driven by physiological improvements. Roosting inside forests might be advantageous regarding the shorter distance for foraging, while bats roosting in caves have to cover longer distances to get access to foraging spots. Generally, there is a high density of roosts opportunities inside forests for tree-roosting bats and the density of bats roosting in forest-microhabitats is smaller compared to colonies in caves (Cardiff & Jenkins, 2016). This might be advantageous, since bats roosting in trees tend to switch their roost frequently when alternative roost sites are available, while this is rare for bats roosting in caves. Roost fidelity from bats in caves might be adversely regarding the increased susceptibility to ectoparasites and, particularly, predation (Lewis, 1995). Bats inhabiting western and south-western Madagascar are not only at risk of predation from animals like Madagascar's largest carnivorous mammal, the endemic Fossa *Cryptoprocta ferox* (when roosts are easily accessible from the ground), the bat hawk *Macheiramphus alcinus* or snakes (e.g. boas) (Cardiff & Jenkins, 2016; Andriafidison et al., 2006a), but are especially endangered by humans (Mickleburgh et al., 2009). Bats are predictable prey for predators and humans when they are roosting at the same sites and caves continuously. Besides, during the day (the inactive phase) bats roosting in trees are often not that accessible to predators, like the bats hanging on cave ceilings, since they are hiding in hollow trees, exfoliated barks, tree crevices or up high in the canopy (Mildenstein et al., 2016). This is particularly true for humans, who hunt bats mostly during the day when bats are resting inside the caves or by equipping the restricted corridors near the cave entrances with fence-like barriers to trap large numbers of bats leaving the caves at sunset for foraging. Since bats inhabiting forests mostly roost solitary or in small groups to avoid overheating and resulting EWL (Speakman & Thomas, 2003), and given the wider distribution across the forest, the hunters' success rate in forests is smaller, compared to hunting large colonies inside or near cave entrances (Goodman, 2006). The most widespread reason for bat hunting is by far for consuming it as bushmeat, as it provides local communities with an alternative source of protein during seasons of food shortage (Mildenstein et al., 2016; Jenkins & Racey, 2008). The most commonly targeted species is the larger Commerson's leaf-nosed bat (*Macronycteris commersoni*) (Cardiff & Jenkins, 2016), but *Triaenops menamena* is also hunted in the western and south-western regions of Madagascar. In the region of the Mahafaly Plateau in south-west Madagascar, an estimated 70.000 – 140.000 microchiropteran bats are trapped and collected each year (Goodman, 2006). Even though *T. menamena* is one of the smaller bat species and is presumably not the main target for bushmeat, it is highly affected by hunting pressure. In caves, bats often form large multispecies assemblages, just like in the examined caves Andranolovy, where the *T. menamena* colony shares a cave chamber with *Macronycteris commersoni* and *Paratriaenops furculus* colonies. Hunting disturbance can not only cause injuries and an increased infant mortality (when pups fall from fleeing mothers), but also disrupt the resting phase causing higher stress levels which affects the overall oxygen consumption and therefore the physiology and energy conservation (Van der Aa et al., 2006). The impacts of hunting pressure elucidate the importance of the bats' roost selection and physiological flexibility, in case they have to switch to other roost sites when disturbances through hunting become too high (MacKinnon et al., 2003).

Therefore, the roost choice is of paramount importance for the bats' energy conservation, but they also exhibit other strategies to minimize their energy expenditure even further. The results display an intraspecific variation in expressed torpor patterns, which represents the physiological flexibility within this species. Besides the most common daily torpor, which is characterized by entering a hypometabolic state accompanied by decreasing skin temperatures, the pattern of "hot" torpor occurred as well, mostly in Kirindy forest. In previous studies, it was also described as "hyperthermic daily torpor", in which skin temperature is high and sometimes even exceed ambient temperatures (Lovegrove et al., 2014). The metabolic rate during these periods of extreme hyperthermia was exceptionally low at 0.06 ml $O_2 h^{-1} g^{-1}$, while skin temperature (35.2°C) was beyond ambient temperature (31°C). This result is comparable to highly reduced TMR of other animals, like the non-lemur primate *Galago moholi*, that expressed a very low oxygen consumption of only 0.09 ml $O_2 h^{-1} g^{-1}$ during torpor (Nowack et al., 2010). However, this moholi bushbaby showed the low metabolic rates while having a skin temperature of around 22°C, which disclose that *T. menamena* is able to express hyperthermia and reduce its TMR to the same level as other animals during hypothermic torpid states. This would be only possible, when the bats relinquish their thermoregulation and reduce the evaporative water loss. In the forest, the relative humidity is decreasing with increasing temperature in the afternoon, at which time bats express daily torpor at high $T_{skin}$ to reduce their energy expenditure, since EWL is increasing at lower RH (Ben-Hamo et al., 2013; Speakman & Thomas, 2003). This strategy of hyperthermic daily torpor is also used by other tropical mammals, like the Philippine tarsier *Carlito syrichta* or the greater hedgehog tenrec *Setifer setosus* (Levesque et al., 2010), and was already shown for the Malagasy bats species *Macronycteris commersoni* (Reher et al., 2018).

The highest torpid $T_{skin}$ during a torpid state was recorded in Andranolovy, from a bat expressing a hot torpor pattern with short torpor bouts. The pattern with short bouts of only 10 to 30 minutes were only recorded in this well buffered cave. Energy expenditure for these periodical arousals might be low overall, since the bats' thermoregulatory costs are low owing to the hot cave surrounding. Furthermore, hypometabolism has its negative affects like a weakened immune function, memory loss and reduced responsiveness to the environment, which could be a reason for this trade-off between costs and benefits from short torpor bouts (Bouma et al., 2010; Millesi et al., 2001). During shorter torpor bouts, the bats body temperature cannot be reduced as much as in long-lasting torpid states. However, the reduction of their metabolic rate and thus energy expenditure was substantial, so that even short torpor bouts (< 30 min) provide considerable energy savings, especially for small bats (Geiser & Brigham, 2000; Turbill, 2006). Even though ambient conditions in Andranolovy cave are favorable, daily torpor was used to reduce the overall diurnal energy expenditure to up to 50 – 90%, compared to days when no torpor is used. Nevertheless, it appeared to be beneficial to "wake up" regularly to reduce the negative effects of torpor, especially since arousal costs within the TNZ are low. For small bats in other microhabitats, it is unlikely to arouse from torpor before sunset when ambient temperatures are below the bats' TNZ (Turbill et al., 2003). That is, the short torpor bouts were only performed by bats in Andranolovy, while bats in Vintany cave or Kirindy forest maintained in a torpid state until sunset.

Overall, the highest occurrence and longest durations of daily torpor, along with the greatest reduction in metabolic rate and skin temperature were expressed by bats in Kirindy forest and all decreased along the fluctuation gradient to the more buffered habitats. This confirms the hypothesis of longer torpor durations in less buffered habitats. Yet, not only the environmental conditions are decisive for the bats' overall metabolic rate and its reductions. There was an effect and a correlation between the bats' body condition and its metabolic rate and skin temperature, which emphasizes the individually differing physiological constraints and the respective responses. The DREE was almost the same in Andranolovy cave and Kirindy forest, which indicates that the lowered costs for bats being in a normothermic state within their TNZ in the cave compensates the increased occurrence of daily torpor and the greater metabolic suppression in the forest. Thus, the first hypothesis cannot be confirmed with this study, since the bats in different populations are highly adapted to its specific microhabitat and use different approaches to minimize their daily energy expenditure. Although, it should be noted that the lowered energetic costs in Andranolovy at night might not represent "the regular scenario", since bats would be flying and foraging during the active phase. Therefore, nightly costs would be higher and since the energy expenditure during the day is higher in Andranolovy anyway, bats in Kirindy forest might spend less energy in general after all. The adaptive potential confirms the third hypothesis, the torpor patterns differed between each site and showed highly individual variations. *T. menamena* expressed intraspecific flexibility regarding different torpor patterns to reduce energetic costs, yet this variability in its physiology was linked to the specific microhabitat. Individuals from the same population showed the same adaptive strategies, while these strategies differed between the roosting sites.

This study revealed new information about the highly adaptive flexibility in the physiology of *Triaenops menamena*. Environments with great fluctuations in temperature and humidity are indeed the most demanding roost sites for this small bat species, yet they are able to cope with it by using different strategies of thermoregulation and hypometabolism. For future prospects, this study indicates that *T. menamena* is capable of contending climatic and environmental changes and responding rapidly to unfavorable conditions with its great adaptive potential. Further studies about seasonal variations are required to understand its broader scope of annual physiological flexibility. Despite the thermoregulatory flexibility, results of the daily energy expenditure and comparisons of advantages and disadvantages of different roost sites, it seems to be more beneficial for *T. menamena* to roost in caves. Therefore, further studies are important to learn more about cave characteristics and to help getting a higher priority in potential conservation plans for these specific refuges. Currently, the greatest threat in Madagascar for this (and many other) species is the loss of habitat and its degradation (Myers et al., 2000). Government projects like the "Durban Vision" and the Madagascar Action Plan "MAP" with the aim to expand protected areas and a sustained use of resources were already good ways to address the problems (Fleischhauer et al., 2008). However, widespread practical implementation was limited due to political unrest and insufficient communication with local communities (Virah-Sawmy et al., 2014). To ensure a long-term persistence of endangered animals in Madagascar, it is of outmost importance to develop more elaborated conservation plans and assure its implementation. Moreover, there needs to be more invested in educational programs for local communities, communicating animal conservation and sustainable land use, as to obviate additional threats like hunting for bush meat.

# Eidesstattliche Erklärung

Hiermit erkläre ich an Eides statt, dass die vorliegende Arbeit von mir selbständig verfasst wurde und ich keine anderen als die angegebenen Hilfsmittel – insbesondere keine im Quellenverzeichnis nicht benannten Internet–Quellen – benutzt habe und die Arbeit von mir vorher nicht einem anderen Prüfungsverfahren eingereicht wurde. Die eingereichte schriftliche Fassung entspricht der auf dem elektronischen Speichermedium. Ich bin damit einverstanden, dass die Masterarbeit veröffentlicht wird.

Ort, Datum					Unterschrift

# Declaration in Lieu of Oath

I hereby confirm that I have written the accompanying thesis by myself, without contributions from any sources other than those cited in the text.
This thesis, in same or similar form, has not been available to any audit authority yet. The submitted written version corresponds to that on the electronic storage device. I agree that the master thesis will be published.

Place, Date					Signature

# Danksagung